\documentclass[sigconf, natbib]{acmart}
\usepackage[utf8]{inputenc}
\usepackage{graphicx}
\usepackage{float}
\usepackage{framed}
\usepackage{soul}
\usepackage{tabularx}
\usepackage[british]{babel}
\usepackage{mdframed}
\mdfsetup{skipabove=8pt,skipbelow=8pt}

\begin{document}

\title{Personal Data Gentrification}
\acmBooktitle{Communications of the ACM}

\author{Juan Luis Herrera, Javier Berrocal, Jose Garcia-Alonso, Juan Manuel Murillo}

\affiliation{%
\department{Department of Computer Science and Communications Engineering}
\institution{University of Extremadura}
\city{Cáceres}
\country{Spain}
}
\email{jlherrerag@unex.es, jberolm@unex.es,} 
\email{jgaralo@unex.es, juanmamu@unex.es}

\author{Hsiao-Yuan Chen, Christine Julien}

\affiliation{
\department{Department of Electrical and Computer Engineering}
\institution{University of Texas at Austin}
\city{Austin}
\state{TX}
\country{USA}
}
\email{littlecircle0730@utexas.edu,}
\email{c.julien@utexas.edu} 

\author{Niko M\"akitalo, Tommi Mikkonen}

\affiliation{
\department{Department of Computer Science}
\institution{University of Helsinki}
\city{Helsinki}
\country{Finland}
}
\email{niko.makitalo@helsinki.fi,}
\email{tommi.mikkonen@helsinki.fi} 
\date{November 2020}

\ccsdesc[500]{Security and privacy~Privacy protections}
\ccsdesc[500]{Security and privacy~Economics of security and privacy}

\begin{abstract}
We live in an era in which the most valued services are not paid for in money, but in personal data. Every day, service providers collect the personal information of billions of individuals, information that sustain their infrastructure by marketing profiles labeled with this information to personal data consumers, such as advertisers. Not all uses of this personal data are for marketing; data consumers can also include, for instance, public health authorities tracking pandemics. In either case, individuals have undergone a process of {\em Personal Data Gentrification}, as data ownership has shifted from individuals to service providers and data consumers, as if the data is worth nothing to the individuals; these new owners then harness the data to obtain large profits. Current privacy-enhancing technologies are beginning to allow individuals to control and share less information. However, not sharing individuals' personal information at all could lead to {\em Personal Data Blight}, in which the potential of personal data in applications that benefit all of society remains forever latent. In this paper, we propose {\em Personal Data Enfranchisement} as a middle ground, empowering individuals to control the sharing of their personal information to shift the business flows of personal information. Based on these insights, we propose a model to gradually and incrementally make a shift from our current situation towards one of Personal Data Enfranchisement. Finally, we present a roadmap and some challenges towards achieving this bold vision.
\end{abstract}

\keywords{Privacy, personal data, data enfranchisement, data gentrification, data blight, data management}

\maketitle

\section{Introduction}

The current landscape of digital services is characterized by a massive circulation of personal information. Every month, over 2 billion individuals share their personal data on platforms such as Facebook, YouTube or WhatsApp~\cite{Hootsuite2020}. Platforms like these are often offered for free; they are mainly sustained by a business model based on marketing the personal data of the individuals who use them~\cite{Elvy2017}. In general, such a business model consists of collecting personal data from individuals to then sell access to labeled profiles from said individuals for purposes such as focused advertising~\cite{CACMFacebook2021}.

The huge digital footprint defined by individuals' personal data, together with available information processing tools, brings unprecedented benefits in terms of knowledge about human behaviors to fuel services that benefit society as a whole. However, certain interests prevent the use of this data at its full potential. In fact, the primary use of these vast stores of personal data is by these same large companies collecting them, for their own economic gain.

We arrived at this situation by experiencing a process of change in personal data usage, one we label {\em Personal Data Gentrification}, since it is analogous to urban gentrification. Personal data is being gentrified, since individuals and the benefits they can obtain with the data are often in the background to cater to large service providers and data consumers, who are wealthier in monetary terms, as well as in technological infrastructure. Maintaining the current approach, in which the huge digital footprints are used almost exclusively for profit, instead of exploiting their potential for societal benefits, makes the digital footprint lose its essence as personal and belonging to individuals. In the same way, maintaining a neighborhood exclusively for profit, instead of exploiting its potential, deviates it from its essence as a community that exists for the benefit of its inhabitants. This approach to the consumption of personal information is leading to {\em catfishing}~\cite{reichart2017follow}, fake profiles in social media~\cite{reichart2017follow}, or fictional {\em influencers} for advertising~\cite{Lee2020}, resulting in larger and less truthful digital footprints.

Personal Data Gentrification is presently evolving to a new stage, in which the potential of data to benefit society becomes latent and the data is instead used for purposes that are outside of an individual's or even society's general interest. This new phase creates situations such as companies creating psychographic profiles of individuals and carefully choosing their advertising so they take certain political actions that are not in their best interest~\cite{Rosenberg2018}. Real-life cases like the Cambridge Analytica scandal make individuals slowly start to be aware of the potential of their collected data and even that data misuse may occur~\cite{Hootsuite2020}. However, it is very complicated for individuals to avoid such practices. In our current, gentrified situation, an individual's data is now the market's property. Individuals are hardly able to know what their data is used for let alone decide how  it is used. However, the recent change in terms of use and privacy policy by WhatsApp~\cite{wup-change} and the following exodus to other similar apps, in particular Signal~\cite{signal-change}, show that there is a growing awareness among the individuals regarding the terms they are ready to accept.

Some proposals to improve this situation involve preventing the gathering of personal information, as is done in the Tor Browser, equipped to prevent any kind of information collection.\footnote{\url{https://torproject.org/}} However, these approaches only shift towards {\em Personal Data Blight}, which completely removes any potential benefits of personal data usage. It is complicated for individuals to benefit from their own digital footprints when they are gentrified; and it is just as complicated when their data is blighted, because the digital footprint becomes very scarce. Our objective is to maintain the circulation of personal information, since we think these footprints should continue to grow, but they should become available for societal benefits and under the control of the individuals themselves. For instance, by asking individuals to share a minimal amount of personal information, Google and Apple were able to create exposure notifications, to assist in mitigating the COVID-19 pandemic.\footnote{\url{https://www.google.com/covid19/exposurenotifications/}}

This work proposes a different point of view to revert the gentrified situation. It assumes the hypothesis that, in the same way that gentrification has been favored by a lack of transparency in personal data collection, processing and use by service providers, reverting gentrification involves bringing light to this process. Thus, as a first step, we propose a model to measure the level of commitment each service provider has to allow individuals to participate in the decision of what is done with their data. A six-level model is detailed in which each service provider can freely measure itself and publish their result. The level a service provider is on provides clear and simple information to the individual about what they can expect in terms of personal data use. The final objective is that the existence of the model, along with market competence, will make providers wish to measure themselves, to obtain the highest level possible and to publish their results. This will naturally change the situation from gentrification to {\em Personal Data Enfranchisement}, in contrast to how gentrification disenfranchises individuals as data owners. The distinctive characteristic of our proposal is that, while opening the data to more purposes aside from service provider profit, it does not go against the business model of service providers; in fact, it benefits from such business creating larger and more useful digital footprints.

In this paper we propose:
\begin{itemize}
    \item A definition and motivation behind the terms {\em Personal Data Gentrification}, {\em Personal Data Blight}, and {\em Personal Data Enfranchisement}.
    \item Identification of the requirements to change the current scenario of personal information.
    \item A definition of a model for enfranchisement to measure the social commitment of a given service provider.
    \item A research agenda towards Personal Data Enfranchisement.
\end{itemize}

The rest of this paper is organized as follows. Section~\ref{sec:motivation} further motivates how the Personal Data has been gentrified and, also, defines the terms {\em Personal Data Blight}, and {\em Personal Data Enfranchisement}. Section~\ref{sec:three} presents a 
model for enfranchisement seeking to increase the commitment with individuals and society, while Section~\ref{sec:emthemes} presents a roadmap that can lead society in this direction. Finally, Section~\ref{sec:conclusion} concludes the paper.
\section{Gentrification of Personal Data}\label{sec:motivation}

In this section, we introduce the concepts of
Personal Data Gentrification, Blight and Enfranchisement. In the process, we introduce and motivate the set of key requirements toward a future of Personal Data Enfranchisement (Table~\ref{tab:requirements}). Throughout the rest of this paper, the underlined words highlight these key concepts and allow the reader to easily link each of the different references to the concept. In addition, we consider three main entities in the personal data market: (i)~data consumers, who are entities willing to access and use labeled profiles; (ii)~service providers, who offer platforms and other services, collect data, build the labeled profiles and market access to said profiles; and (iii)~individuals, who make use of the platforms and services and have their data collected.

The main motivation behind this paper is to address the lack of awareness by the general public about Personal Data Gentrification, as well as its future consequences if this gentrification is prolonged in time. Currently, individuals live happily, seeing how they get services and social media free of charge. The truth, however, is that most of these applications do have a price, which is paid for with personal data. Furthermore, even when individuals do pay for services with money, they have a perception that the provider takes more care with their personal data, but this is rarely the case~\cite{RedShell2018}.

One of the keys to Personal Data Gentrification is the pervasive End User License Agreement (EULA). A EULA is an agreement that the individual needs to sign in order to obtain access to a service, while also agreeing to have their data collected. A EULA includes information such as what pieces of personal data are obtained from individuals (not including those that are inferred) and under which conditions they are collected. 
By agreeing to this EULA, individuals expect service providers to be guardians of their personal information, only using it in the ways the EULA allows. The truth is that a very scarce minority of individuals read these EULAs~\cite{obar2020biggest}, and, further, the use of {\em dark patterns} is commonly used to get individuals to accept EULAs, even if the conditions are outside their best interest~\cite{Gray2018}; as a result EULAs are rarely transparent to individuals.

If sharing personal data can be of benefit in principle, it can also become a problem when the transferred data is used for purposes that go against its owners, a situation that appears when some data consumers perform activities with information from unwilling individuals, in what is known as {\em personal data misuse}. This is a huge problem even today, despite previous attention -- research conducted in 2020 shows that data misuse, such as by ransomware or spam, still exists in Facebook even after the Cambridge Analytica scandal~\cite{Farooqi2020}.

Another key feature of the EULAs that contribute to Personal Data Gentrification is the fact that they present individuals with a {\em take it or leave it} situation. This is interesting, because, of all three of the parties involved in the process, only individuals are burdened with such a lack of negotiation. Data consumers can negotiate with service providers about the pricing of the data they consume (e.g., advertisers may pay different rates depending on how many individuals they want to receive their advertisements), as well as on which data they consume, why and how. In fact, the European General Data Protection Regulation (GDPR) defines the Data Processing Agreement (DPA), which must be signed by both service providers and data consumers, so both can negotiate how the consumption is performed~\cite{GDPR}. Furthermore, full control of the agreement rests with the service provider, who is always responsible for enforcing it. The situation is very different in the relationship between an individual and a service provider: the individual either accepts the EULA, thereby allowing the service provider to take all the data specified in it and fully trusting the service provider to uphold it, or refuses to accept the EULA and is unable to use the service at all.

Regulated by EULAs, service providers build their main asset: a set of labeled profiles from the collection of personal data along with new information inferred through the use of data mining~\cite{CACMFacebook2021}. Because these profiles are their primary asset, service providers desire to protect this data, rather than sharing it with their competitors. So they store and protect these profiles in their own facilities, making them incoherent between different providers and unavailable for societal benefits. In this way the individual also lacks ownership of their own personal data, as they do not have control over its storage. Thus the process for Personal Data Gentrification is closed.

The above leads us to the following Personal Data Gentrification scenario: individuals are faced with a {\em take it or leave it} situation and are subject to EULAs, while they fuel the services they seemingly consume for free by having their personal information collected and potentially misused; service providers privately store data and data consumers have to settle with muddled profiles, as different service providers may have different profiles of the same individual because they collect different pieces of information. The development of systems that make use of personal information is driven by profit, and therefore only well-resourced entities are able to develop them, making it complicated to develop data uses that focus on benefiting society. This is especially dangerous when individuals' preferences regarding their personal information usage are skipped as long as the consumer pays~\cite{Farooqi2020}, because it opens the door to data misuse.

Therefore, we define Personal Data Gentrification as a process throughout which individuals' digital footprints become unavailable for societal benefits and lose their essence as personal and belonging to individuals. Analogously, Personal Data Blight is a process that minimizes the size of digital footprints, thus leaving its potential unavailable for all parties. Personal Data Enfranchisement is a process that makes digital footprints become available for societal benefits and regain their personal essence. The goal of our paper is to lay out a model that allows us to measure a service provider's commitment towards Personal Data Enfranchisement, avoiding a gentrified future in which we find that an increase in the flow of personal data brings only an increase in profits and not in societal benefits, as well as avoiding a blighted future, curtailing the sharing of personal information, and devoid of any use of personal digital footprints. With all the above in mind we enumerate the requirements for Personal Data Enfranchisement in Table~\ref{tab:requirements}.

\begin{table}[htb]
\small
\caption{Requirements for Personal Data Enfranchisement}
\begin{tabularx}{\columnwidth}{|c|X|}
\hline
{\bf Keyword} & {\bf Requirement description} \\ \hline
\ul{Awareness} & Individuals must be aware of the value of their personal data and the fact it is being collected and shared. \\ \hline
\ul{Transparency} & Service providers must be transparent to individuals about the data they collect and who can access it. \\ \hline
\ul{Purpose} & Data consumers must report their purpose for the usage of personal data to service providers. Service providers must also be transparent about this information with individuals. \\ \hline
\ul{Negotiation} & Individuals, service providers and data consumers must all be able to negotiate and reach an agreement on the acceptable usage of personal data. \\ \hline
\ul{Trust} & Individuals must be allowed to deposit their trust on service providers or data consumers, as well as to revoke the trust if they feel their interests have been harmed. \\ \hline
\ul{Coherence} & Data consumers must have access to coherently labelled profiles throughout different service providers to increase their usefulness. \\ \hline
\ul{Availability} & Personal data must be available for the development of services that are beneficial to society if individuals wish it to be. \\ \hline
\end{tabularx}
\label{tab:requirements}
\end{table}
\section{A Model for Enfranchisement}\label{sec:three}

To enable the transition from Personal Data Gentrification to Personal Data Enfranchisement, the business flows of data must shift towards a balance between an individual-centric model and the current profit-centric one. In an enfranchised, the impetus behind data usage is not only the profit obtained, but also the benefits the usage brings to the data owner and to society.

\begin{figure*}
    \centering
    \includegraphics[width=0.75\linewidth]{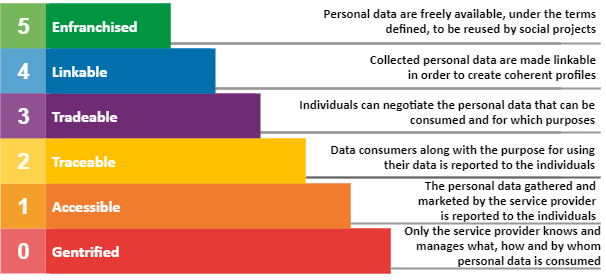}
    \caption{Model for Personal Data Enfranchisement.}
    \label{fig:PDE_Qualification_model}
\end{figure*}

Meeting the requirements presented in the above section is very difficult to envision, given the current scenario. A wholesale replacement of the current system is not feasible. The shift must be made as a set of incremental changes to the information and business processes of current entities. To catalyze this transformation, this section presents a Model for Personal Data Enfranchisement (Figure~\ref{fig:PDE_Qualification_model}). This model consists of six levels, each of them implying a particular change in how personal data is managed to increase the commitment to individuals and society. These levels maintain the role of each entity untouched, while incrementally empowering individuals, increasing the social benefit and balancing the trade-offs that appear as a result, so that all entities can obtain benefit from this transformation.

\begin{mdframed}
\section*{SIDEBAR: Personal Data Enfranchisement: A Historical Perspective}

Personal Data Gentrification has become increasingly significant only in the last few years. In 2005, social media users were still a minority and the personal information individuals shared on the Internet was negligible~\cite{Perrin2015}. Individuals' digital footprint was scarce---so scarce, in fact, that data misuse was very rare. Data consumers were often limited to the service providers themselves, who consumed personal data to improve their own services, such as adding recommendations~\cite{Adomavicius2005}.

From the mid 2000s, individuals’ digital footprints have grown at an unstoppable pace, fostered predominantly by social networks~\cite{Samur2018}. This change has shifted the role of individuals, who are more active in sharing and publishing personal content and thereby increasing their digital footprints~\cite{turcotte2011web}. This increase in data circulation fostered a new marketing strategy known as Digital Marketing or Marketing 2.0~\cite{Tiago2014} that uses individuals’ digital footprint to create specific communications channels with targeted individuals in order to better engage them~\cite{tomvse2014marketing}. This has resulted in the shift we depict in Figure~\ref{fig:gentrifuture}.

\begin{center}
    \includegraphics[width=.9\textwidth]{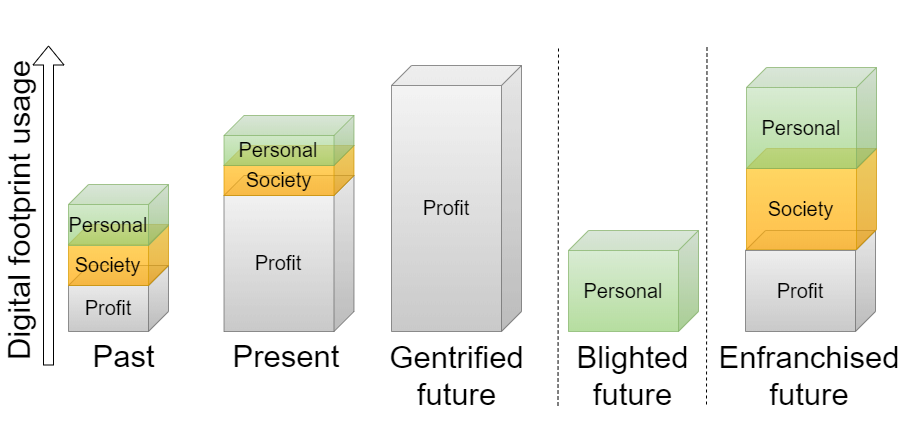}
    \captionof{figure}{Past, present and possible future situations of personal data as digital footprint.}
     \label{fig:gentrifuture}
\end{center}

The increasing digital footprints have also been  driven by the massive deployment of personal Internet-connected mobile devices. Numerous social media applications have gone mobile because they allow individuals to be constantly connected and, thus, share a higher volume of personal content. In addition, sensors in mobile devices such as GPS, microphones or cameras have contributed to increases in the quantity of the data. As a result, mobile marketing~\cite{Kaplan2012} has developed, which allows companies to further customize their messages implementing a one-to-one communication channel~\cite{Kaplan2012}. The use of individuals' data footprints as part of initiatives that benefit society as a whole is still scarce, and these initiatives often come from large companies~\cite{Apple2020}. 

In parallel to these emerging commercial uses of individuals' data, over the past several years, various platforms have started to emerge that increase individuals' awareness of and control over their own personal data. One of the earliest of these, MyLife Digital provides individual consent management frameworks that are more flexible than the plain {\em take it or leave it} approach.\footnote{\url{https://mylifedigital.co.uk}} MyLife Digital's tools also support service providers in ensuring compliance with individuals' consents related to data use. Providing an even broader umbrella, the international non-profit MyData Global supports collaborations among entities with interests in building a human-centric personal data ecosystem.\footnote{\url{https://mydata.org}} However, many societal and technical challenges remain to be solved; our efforts construct a model for enfranchisement within which to explore dismantling the existing technical barriers.

Others have explored what an individual's role might be in a future data marketplace. BitsAboutMe,\footnote{\url{https://bitsabout.me/}} for instance, allows an individual to harness their digitized personal data and make trades with it, for money. Effectively, the BitsAboutMe model cuts out the service provider, allowing an individual to negotiate directly with the data consumer. DigiMe similarly allows individuals to create a replicated personal store of their digital data from a variety of platforms then selectively share this data only as the individual chooses.\footnote{\url{https://digi.me}}

We refer to this class of emerging tools as {\em data enfranchisement service providers}, in contrast to the more generic {\em service provider}. These data enfranchisement service provider tools have improved the ability for individuals to understand the nature and value of their personal data. To those who opt-in, these services also expose aspects of the processes that are used by service providers to monetize personal data. These tools require individuals to connect their accounts (and thereby their data) from existing service providers to the data enfranchisement service provider. This creates a second, parallel stream from individual to data consumer. This second stream is entirely under the control of the individual, but its existence does not change the existence (and potential misuse) of the primary stream that passes through the traditional service provider. Therefore, these existing tools help to address the requirement for awareness, but they still do not meet the goal of putting fine-grained control of the release and monetization of that data in the hands of individuals.
\end{mdframed}

To help understand the effects of each level, we use a running example centered on the usage of voice assistants, of the likes of Alexa or Siri. This voice assistant collects three kinds of personal information: spoken language, speech patterns and breathing and coughing patterns, that are used by the service provider to offer different functionalities. This information is also used to label the individuals' profiles to enable targeted marketing for data consumers. In our case, we consider a data consumer whose purpose is offering relevant advertisements to individuals based on their labels; and a research initiative, whose purpose is creating an application able to detect COVID-19 from voice patterns.\footnote{These applications are not merely a thought experiment, as there are researchers investigating the diagnosis of COVID from voice patterns~\cite{Brown2020}; and there are also systems able to provide personalized advertisement based on voice data analysis~\cite{freeman2010system}}

{\bf Level 0 - Gentrified.}
This is the level at which most service providers would be qualified today. The service provider gathers the personal data required to develop and offer their services. This information is then used to create the individuals' labeled profiles and offer them to data consumers. Individuals’ personal data is gentrified; they are \ul{unaware} of what information is gathered and marketed, and they can not use it for their own benefit.

In our voice assistant example, an individual gives all their personal information to the service provider, and individuals do not have the means to negotiate over their data since the EULA forces them to accept a {\em take it or leave it} service offering. Service providers and data consumers do negotiate with each other over the use of individuals' personal data, using DPAs. With these DPAs, the advertiser shows their advertisements depending on individuals' spoken language and speech patterns, while the COVID initiative focuses on coughing and breathing patterns. Individuals do not know what information is gathered, who consumes that information or what are the purposes behind this consumption.

At level 0, the service provider offers functionalities in exchange for personal data. Nevertheless, as long as the individuals are not aware of the volume, the value of their personal information and the current, gentrified situation, this situation cannot provide benefit to anyone other than a paying service provider. Only with awareness, a fair playing field for all of the involved entities can be implemented in order to bring benefits to the whole society.

{\bf Level 1 - Accessible.}
This level brings \ul{awareness} to individuals about the personal data that is gathered and that is exposed to data consumers. All the information individuals are given about the usage of their personal information is in the EULA, which provides only broad strokes and is not accessible to individuals~\cite{obar2020biggest}. Only service providers keep a fine-grained trace of what information is obtained from their individuals, for what and which labels are exposed to data consumers.

At level 1, the service provider increases its \ul{transparency}, communicating two types of information with individuals: what pieces of their personal data are being gathered and which ones are offered as labels to data consumers. Different mechanisms are also implemented so that these registers are easily accessible, understandable and inspectable directly by individuals. In moving from level 0 to level 1 in the running example, the individual becomes aware that their voice, breathing and coughing patterns, as well as spoken language, are collected, and that the profile generated with all of these data points is labeled for data consumers with their characteristics.

This change primarily benefits individuals, who are now able to choose whether to use a service or not based on their own perception of trustworthy entities or their own data sensitivity. However, service providers can also benefit from the increased communicativeness: individuals can select those service providers that they deem more trustworthy, such as those that collect less data (for instance, in our running example, only the language individuals speak). These providers can use the provision of increased awareness as a potential selling point.

{\bf Level 2 - Traceable.}
Level 2 focuses on increasing the transparency of the data consumption ecosystem. With level 1, individuals are left to wonder which data consumers are using the personal data exposed by service providers. In the running example, individuals do not know that the advertiser is interested in the spoken language and the speech patterns; whilst the COVID researchers are focusing on the breathing and coughing patterns. In addition, the reasons why each data consumer is interested in those labels is totally unknown to individuals, which leaves them totally blind about how and for what purposes their personal information is used. In our example, individuals do not know if the advertiser wants their profiles to contribute to research, or if the researchers want to profile them by language. 

At level 2, the service provider fully implements the concept of \ul{transparency} and also integrate the \ul{purpose} as part of this transparency. This is done first, by communicating the identities of the data consumers and which labels they make use of; and, second, by obtaining the purposes for the data consumption and reporting them to the individuals. For the voice assistant example, the service provider will report that the advertiser is interested in the spoken language and  speech patterns in order to send customized advertisements to the individual, whilst the researchers are using the breathing and coughing patterns for detecting COVID-19. 

Level 2 brings benefits similar to those of level 1: at this level, the entire personal data management ecosystem becomes transparent to individuals, and service providers affiliated with data consumers who have purposes that individuals find beneficial, such as COVID research in our example, can use this information as a selling point.

{\bf Level 3 - Tradeable.}
Level 3 focuses on allowing individuals to negotiate what labels can be used by data consumers, by whom and for what purposes. After level 2, individuals have knowledge about the manner and purpose of the consumption of their profile, but no real control over it, since the agreement individuals need to accept with the service provider is still only a {\em take it or leave it} EULA. In our running example, individuals cannot allow their breathing and coughing patterns to be used in COVID research without allowing their language and speech patterns to be used for advertising as well. They are also unable to limit the advertiser's knowledge of their spoken language, or to allow researchers only to focus on their coughing patterns.

At level 3, the service provider implements and offers different processes for empowering individuals to \ul{negotiate} which specific labels they allow to be marketed about them, for which purposes each label can be consumed, and who can consume it. There is already a mechanism that regulates these conditions: the DPA that service providers and data consumers agree on. These DPAs can be extended to individuals, thereby having all three parties -- individuals, service providers and data consumers -- negotiate in the establishment of DPAs. In addition, at this level, the service provider has these mechanisms adapted to be easily accessible and understandable by individuals or to be delegated to third-party entities \ul{trusted} by them.
Applying these negotiating processes in the voice assistant running example, some individuals may decide to restrict advertisers to only use labels about the spoken language, while researchers may be allowed to make use of labels about their coughing, breathing and even speaking patterns to perform research.

Aside from the benefits for individuals that this level brings, and although some data consumers may see a reduction on the amount or quality of the data they can consume, service providers and data consumers obtain a potential increase in their individual base, as those individuals who rejected using the service because of a few purposes of use may now opt in, given the ability to exert more individual control.

{\bf Level 4 - Linkable.}
Level 4 is focused on increasing the \ul{coherence} of the personal information gathered by service providers. Up to this point, each service provider privately stores all of the data they take from individuals, discouraging data reuse. Thus, each service provider constructs different profiles for the same individual. This situation may lead to inconsistencies. Different service providers may have contradictory information associated with the profiles and labels for the same individual. This should not be surprising since different service providers collect information focusing on different purposes and also because humans tend to have different behaviors depending on each service provider's focus. In the running example, the individual can have another voice assistant from a different provider for helping them to control different actions in their car; both assistants gather the same information, but each provider will only have a partial view of the individual and can label them differently (as quiet, calm or nervous) because they behave differently at home and driving their car. 

The ideal situation is to take advantage of the complete digital footprint that each individual leaves on the Internet. To that end, all the labels and profiles collected by the different service providers should be combined. Thus, the inconsistencies can be eliminated and, in addition, different viewpoints can be created for the same individual from different facets of their lives even when they can be contradictory. Thus, at level 4, the service provider makes their computed labels and profiles available and linkable with the profiles of the same individuals by other service providers. This would require, first, reference mechanisms to uniquely identify each individual and, second, for the service provider to use them to identify the profiles they manage. These identifiers could be simple and globally unique, although they can also be complex or privately created in an ad-hoc manner and shared by groups of service providers who wish to allow the linkage of the profiles they each handle. The potential linkage of the data managed by a specific service provider would be given by the universality of the mechanism it uses. The compliance degree of level 4 would be determined by coverage of the profile identification mechanisms they are based on.

Of course, who is linking each profile is also reported to the data owners. In addition, processes are implemented by the service provider to validate which entities can link the stored data according to the DPAs signed in the previous level. Then, when another entity requests any of these profiles, a reference to access it is provided, similar to the behavior with linked data. In our example, the information gathered by both providers is exposed in a linkable format so that any entity can create a more coherent profile of the individuals pointing to the labels obtained by the voice assistants in both the individual’s house and car. In addition, it is easier for other providers to know if they can reuse some specific label already set.

The benefits of level 4 are mainly for service providers, who can store cleaner information per individual, and for data consumers, who regain some data quality. In addition, individuals can see the opportunity for all the data they leave on the Internet to be used together, and society in general can start to generate rich and complete knowledge bases of humanity.

{\bf Level 5 - Enfranchised.}
The objective of level 5 is for service providers to increase their social commitment by providing \ul{availability} of the labeled profiles to social projects that are beneficial to the individuals. Note that while level 4 implies that data is made linkable, service providers can still keep it for exclusive use or for linking in networks of service providers that only seek mutual benefit.

At level 5, the service provider opens individuals' profiles to be consumed by social projects benefiting society. For that, it implements mechanisms to grant access to the profiles to public or private initiatives focused on the general benefit of society. Of course, all the grants provided must be compliant with the preferences and permissions of the profiles' owners. Also, the precedence of the initiatives should be well identified in both their motivations and their aims. Going back to the example, a new research initiative could appear with the aim of building learning models to detect depression from speech patterns~\cite{Roniotis2018}. They could ask for access to the already labeled profile for the COVID initiative if the DPA allows it. Mutual sharing would boost both initiatives.

The final benefit is therefore a general \ul{availability} of personal information for contributing not only service providers but also the societal purposes pursued by individuals, allowing these initiatives to truly and fully extract the potential of our digital footprint, instead of wasting it. In addition, it increases the \ul{trust} of individuals on the personal data management ecosystem.

Personal Data Enfranchisement, when reached by following the presented model, is able to change information flows towards a balance between an individual-centric model and a profit-centric one without needing to dramatically change the roles or behaviour of any of the three involved entities. The different changes that it proposes to the information flows bring benefits for all, individuals, service providers and data consumers, democratizing personal data usage.

\section{A Roadmap to Enfranchisement}\label{sec:emthemes}

The model for enfranchisement described in the previous section is a proposal that tries to establish a logical path from gentrification to enfranchisement. However, within the scope of this article only a brief description of each level is provided indicating what characteristics it adds with respect to lower levels. The path to enfranchisement will require not only a complete development of the model but also changes in the current situation at various levels ranging from technical to social, political and regulatory aspects.

The development of the model for enfranchisement will require a detailed description of each level establishing, in an unambiguous way, what is required and what is not supported for each level along with an adequate justification for it. Thus, e.g., it will be required to specify which are the key areas in which the management of personal information is structured, such as negotiation for collection, collection procedures, inference of new information, negotiation for private and commercial use or negotiation for social use. Good practice guidelines will also be required for each of the above areas at each level of the model as well as the identification of inappropriate practices. The specified model must scrupulously comply with all regulations on the use of personal data such as GDPR~\cite{GDPR} or ISO/IEC 27701~\cite{ISOIEC27701}. Although the specification of the model for enfranchisement does not have to be unique, in an ideal situation it would be developed in a joint public and private initiative so that all involved agents can contribute their perspective.

The existence of a specific model for enfranchisement will demand certification procedures so that the service providers can certify their commitment to the management and use of personal information. These certification procedures will not be of much interest initially to established service providers, but they will be for new service providers that enter the market and want to use their certification as a distinctive value. This will encourage established service providers to consider certification as well. In any case, the need for certification will bring new agents into the market whose business is based on both consulting and auditing. An example of this could be certification authorities for personal data use.

It can be expected that the existence of the model for enfranchisement, the certification managed by service providers, the use of the certification as a marketing asset, the emergence of new procedures available to individuals in order to negotiate the use for which their data are transferred will cause society to become more aware of the use and potential of personal data. In this sense, a similar effect can be expected to that which has been achieved by the energy labeling of household appliances. Labels such as Energy Star have become very valuable and individuals are incited to acquire a new household appliance when they have energy certification labels~\cite{banerjee2003eco}.

In any case, the development of the model for enfranchisement will require political momentum. The situation will hardly change if there is no interest on the part of governments to make the societies they lead more aware of and involved in the use made of the data. Raising this model to a regulatory level would undoubtedly be the most effective push to begin to move decisively towards enfranchisement.
In addition, the application of the model for Personal Data Enfranchisement and the certification process have also associated some technical and social emerging themes that are presently surfacing, and that need to be further investigated by the research community and developed by the industry and governments.

{\bf Conveying the technical underpinnings of personal data value, use, and implications.} While this theme is not {\em entirely} technical, it requires a close collaboration between individuals and those with deep technical expertise. An unfathomably wide chasm exists between the individuals who should own their personal data and those who understand the computers and networks behind how that data is created, stored, shared, and used. As an example, consider a ``quiz'' on a social media platform: {\em What food matches your personality?} After answering a series of questions ranging from ``What were you like in high school?'' to ``Choose a photo of your favorite baby animal'', the quiz ``predicts'' what food best matches your personality. Such a quiz seems innocent. But consider that it also contains questions like ``How do you spend your free time?'' and ``What is your style?'', and, with a little bit of imagination, you can start to see how the purveyor of the quiz might be able to monetize the data collected in responses. However, these kinds of techniques are precisely aimed at those who are unaware of the importance of this data. Other similar techniques have been used by ``research'' apps that were later banned from stores~\cite{facebook-app} or dark patterns in cookie policies~\cite{Gray2018}. The movement from level 0 to 1 should not only be an effort from service providers to make individuals aware of the personal data that is gathered. They should also be able to understand the implications of sharing personal data. Therefore, an educational effort is also necessary on the part of educators and technical experts to better transmit how the personal data can be used and its value. These efforts need to be broad and include individuals from the moment they start generating data, such as teenagers or young adults.

{\bf Tools that allow individuals to inspect the use of their personal data.} Emerging efforts by the companies, such as MyLife Digital, MyData Global or DigiMe, empower individuals to control how their personal data is potentially used. A next step is to give individuals visibility into how their labeled profiles are {\em actually} used, whether it is identifiable (i.e., tied to their identity) or in aggregate with other profiles, and to communicate who uses them and what for, key objectives in levels 1 and 2 of our model. Moreover, these tools must be easily understandable and inspectable by these individuals at any moment. This calls for technical solutions, mainly related to user experience, that should be integrated in service providers' systems.

{\bf Negotiation of data usage conditions}. Standard-form license agreements are a nuisance to individuals~\cite{bakos2014does}, and therefore they are often agreed to without considering the terms~\cite{obar2020biggest}. Alternative ways to negotiate data usage conditions are needed, mechanisms that are truly considered by individuals, as stated in level 3 of our model for Personal Data Enfranchisement. Obviously, asking all individuals to define their own contracts is not technically feasible, so certain predefined models, with possibly an option to customize some parts of them, are still needed. These predefined models need not originate only from service providers, but governments or other organizations -- such as MyLife Digital or MyData Global -- could propose recommendations and coordination, in analogy to Open Source Initiative \footnote{\url{https://opensource.org/}} providing information about open source licenses. Moreover, new business models may emerge in which individuals can delegate the negotiation of the data usage to certain organizations (public or private) based on basic guidelines.

{\bf Accountability of data usage}. To a degree, one can argue that by deliberately providing data to a service, there is no question about individuals' consent. However, this view does not withstand closer scrutiny. Over time, things often change, and a consent given previously might no longer be valid -- the company an individual originally agreed with may have been purchased or merged, or may have changed their business approach. Therefore, it is reasonable that there must be a mechanism for revoking consent or for specifying a deadline for consent after which consent is revoked, resulting in removal of the individual's labelled profile. The same mechanism is needed for other cases, such as data misuse.

{\bf Personal profiles coherence and storage}. Another key matter to address in level 4 of our model is the coherence of personal profiles and the storage of these profiles. This is expected to be performed using techniques akin to linked data, requiring new systems that are not only able to work seamlessly with locally stored profile labels and linked ones, but also to properly store and link the label information. Furthermore, these systems need to treat information securely, as distributing the data from global enterprises to smaller companies can also lead to problems, in the form of these entities' capabilities to handle data securely and privately. Related risks were evidenced by a recent case in Finland, where personal mental health data was leaked from a small private company~\cite{vastaamo}. Profile labelling systems that are coherent, interoperable, secure and private are key to the higher levels of Personal Data Enfranchisement, and can reduce the risks of smaller service providers handling sensitive information. We can find some works in literature that partially address some of these challenges. Technical works, such as Solid~\cite{Berners-Lee2020}, Human Microservices~\cite{Laso2020}, Human Data Model~\cite{makitalo2020internet}, Citizen Digital Twin~\cite{CDT2021} or Paco~\cite{Wendt2018}, allow individuals to better control their personal data and to store labels for profiles in a decentralized, linked manner. These technical solutions must be gradually incorporated by the industry to improve the management of personal data.

{\bf Economic and social impact}. The final topic in this research agenda is the economic and social impact of data openness. In levels 4 and 5, profiles are linked akin to the concept of linked open data~\cite{Berners-Lee2006}, but with the addition of solid security requirements. The benefits of this openness for service providers and data consumers are clear: with it, they can obtain a single, coherent personal profile for each individual, as well as save on costs of data storage. However, a main question still remains. Once level 5 is reached, and information is available for society, what are the benefits of this open, linked approach? We believe that the benefits that enfranchised personal data (level 5) would bring to society are very similar to, or can be extrapolated from, the ones that are already obtained by the Open Data business model. Opening the data of the European Union governments has brought many direct and indirect benefits. Considering both the direct and indirect market size of Open Data, the cumulative total market size is estimated between 1,138 and 1,229 bn EUR~\cite{Berends2020}. Furthermore, there are not only economic benefits from opening data, but also societal benefits. For instance, it was estimated that Open Data has the potential of saving 1,425 lives a year~\cite{Berends2020}. With this promise for opening data at government level, a key question that must be answered during the next few years is: what are the direct and indirect benefits that opening personal profiles would provide to society? In this sense, some governments, such as the European Union, are starting to discuss new regulations to strengthen data-sharing mechanisms~\cite{EU2020}.
\section{Conclusion}\label{sec:conclusion}

The personal data of millions of individuals is collected every day, and exploited only for profit, fueling platforms whose business models are based on marketing labeled profiles of individuals. The use of individuals' digital footprints is gentrified, locked behind a paywall, limiting the possibilities for their usage in initiatives that directly benefit the individual or contribute to society. Moreover, the rise of personal data misuse has motivated proposals that improve privacy by preventing the creation of digital footprints, blighting the usage of personal data. We presented Personal Data Enfranchisement, a process to revert both Personal Data Gentrification and Personal Data Blight by allowing digital footprints to regain their personal essence and become available for societal benefit.

The path towards Personal Data Enfranchisement is not only a matter of technical challenges: social, political and regulatory aspects must also be considered. As a first step towards addressing these challenges, we present the Model for Personal Data Enfranchisement, with six levels of incremental changes that allow a service provider to move from Personal Data Gentrification to Personal Data Enfranchisement. Moreover, we present a roadmap to enfranchisement, and discuss some of the key emerging themes and future challenges for society members, politicians, technicians and regulators. In the future, we expect the challenges presented in our roadmap to be addressed, including the promotion of a task force involving both public and private entities, such as governments and companies, for a detailed definition of the model.

\begin{acks}
This work was supported by the projects RTI2018-094591-B-I00 (MCI/AEI/FEDER,UE), the 4IE+ Project (0499-4IE-PLUS-4-E) funded by the Interreg V-A Espa\~na-Portugal (POCTEP) 2014-2020 program, by the Department of Economy, Science and Digital Agenda  of the Government of Extremadura (GR18112, IB18030), by Business Finland (project AIGA: AI Governance and Auditing), by the Academy of Finland (project 328729), by the European Regional Development Fund, and by the National Science Foundation (CNS-1703497). Any opinions, findings, and conclusions or recommendations expressed in this material are those of the author(s) and do not necessarily reflect the views of the sponsors.
\end{acks}

\bibliographystyle{ACM-Reference-Format.bst}
\bibliography{biblio}

\end{document}